\documentclass[11pt]{article}
\usepackage{jheppub}
\usepackage{amsmath}
\usepackage{amssymb}
\usepackage{braket}
\usepackage{bbold}
\usepackage{pifont}
\usepackage{color}

\newcommand{\beqn}{\begin{eqnarray}}
\newcommand{\eeqn}{\end{eqnarray}}
\newcommand{\dd}{\mathrm{d}}
\newcommand{\nn}{\nonumber}
\newcommand{\Tr}{\mathrm{Tr}}

\newcommand{\gmn}{g_{\mu\nu}}

\newcommand{\fmn}{f_{\mu\nu}}

\newcommand{\hmn}{h_{\mu\nu}}

\title{Ghost-free completion of an effective matter coupling in bimetric theory}
\author{Marvin L\"uben \& Angnis~Schmidt-May}
\affiliation{Max-Planck-Institut f\"ur Physik (Werner-Heisenberg-Institut)\\
F\"ohringer Ring 6, 80805 Munich, Germany}
\emailAdd{mlueben@mpp.mpg.de, angnissm@mpp.mpg.de}
\abstract{We consider a particular set of ghost-free interactions for three spin-2 fields in which we 
freeze out the dynamics of the metric tensor that couples to the matter sector.
Integrating out the non-dynamical degrees of freedom  in vacuum results in a
ghost-free bimetric theory. In the presence of the matter source, which we treat as 
a small perturbation, the equations for the non-dynamical field can be solved perturbatively
for its vierbein. 
This results in ghost-free bimetric theory in vierbein formulation with a modified matter coupling. 
To lowest order in matter perturbations, we precisely obtain an effective matter coupling that has been 
suggested earlier in the literature. This coupling contains a linear combination of the two vierbeine 
whose corresponding metric fluctuation coincides with the massless spin-2 mode.
In the past, bimetric theory with this symmetric coupling 
has been treated as an effective theory valid at low energies without a ghost-free
completion. Our results demonstrate that the effective matter coupling
 can be rendered entirely ghost-free by including the higher-order corrections 
 obtained from the trimetric setup. }

\begin{document} 
%%%%%%%%%%%%%%%%%%%%%%%%%%%%%%%%%%%%%%%%%%%%%%%%%%%%%%%%%%%%%%%%%%%%%%%%%%%%%%%%%%%%%%%%%

\begin{flushright}
\hfill{MPP-2018-55} \vspace{20mm}
\end{flushright}
\maketitle
\flushbottom

%%%%%%%%%%%%%%%%%%%%%%%%%%%%%%%%%%%%%%%%%%%%%%%%
\section{Introduction}
%%%%%%%%%%%%%%%%%%%%%%%%%%%%%%%%%%%%%%%%%%%%%%%%

%%%%%%%%%%%%%%%%%%%%%%%%%%%%%%%%%%%%%%%%%%%%%%%%
%\subsection{Motivation \& context}
%%%%%%%%%%%%%%%%%%%%%%%%%%%%%%%%%%%%%%%%%%%%%%%%

Bimetric theory~\cite{Hassan:2011zd} is a model for a massive spin-2 field interacting with a massless one. 
It generalizes both general relativity (GR), which describes a massless spin-2 field,
as well as nonlinear massive gravity~\cite{deRham:2010kj}, which describes a massive 
spin-2 field alone. For reviews of bimetric theory and massive gravity, see~\cite{Schmidt-May:2015vnx} 
and~\cite{Hinterbichler:2011tt, deRham:2014zqa}, respectively. Bimetric theory can be further generalized
to multimetric theory~\cite{Hinterbichler:2012cn} which always includes one massless spin-2 and several 
massive spin-2 degrees of freedom. This is in agreement with the fact that interacting theories for more 
than one massless spin-2 field cannot exist~\cite{Boulanger:2000rq}.

The bi- and multimetric actions are formulated in terms of symmetric rank-2 tensor fields,
whose fluctuations around maximally symmetric backgrounds do not coincide with the 
spin-2 mass eigenstates~\cite{Hassan:2012wr}.
The form of their interactions is strongly constrained by demanding the 
absence of the Boulware-Deser ghost instability~\cite{Boulware:1973my}. 
The ghost also makes it impossible to couple more than one of the 
tensor fields to the same matter sector, at least not through a standard minimal coupling,
mimicking that of GR~\cite{Yamashita:2014fga, deRham:2014naa}.
A consequence of this is that the gravitational force is necessarily mediated by a superposition
of the massless and the massive spin-2 modes and not by a massless field alone, as one might expect.
It is an interesting open question whether more general matter couplings can be realized in
bimetric theory without re-introducing the ghost.

It has been shown that the ghost does not appear at low energies
if one couples the two tensor fields $\gmn$ and $\fmn$ of bimetric theory 
to the same matter source through an ``effective metric" of the form~\cite{deRham:2014naa},
\beqn\label{effmetr}
G_{\mu\nu}=a^2g_{\mu\rho}+2ab\, g_{\mu\rho}\big(\sqrt{g^{-1}f}\,\big)^\rho_{~\nu}+b^2\fmn\,.
\eeqn 
Here, $a$ and $b$ are two arbitrary real constants and the square-root matrix $\sqrt{g^{-1}f}$ is defined via
$\big(\sqrt{g^{-1}f}\,\big)^2=g^{-1}f$.

Ref.~\cite{Noller:2014sta} suggested a similar expression for 
bimetric theory formulated in terms of the vierbeine
$e^a_{~\mu}$ and $v^a_{~\mu}$ in $\gmn=e^a_{~\mu}\eta_{ab}e^b_{~\nu}$ 
and $\fmn=v^a_{~\mu}\eta_{ab}v^b_{~\nu}$. Namely, they couple the metric,
\beqn\label{effmetrvb}
\tilde{G}_{\mu\nu}=\big(ae^a_{~\mu} +bv^a_{~\mu}\big)^\mathrm{T}\eta_{ab}\big(ae^b_{~\nu} +bv^b_{~\nu}\big)\,,
\eeqn
to matter. This metric coincides with (\ref{effmetr}) if and only if the symmetrization condition,
\beqn\label{symcond}
e^a_{~\mu}\eta_{ab} v^b_{~\nu}=v^a_{~\mu}\eta_{ab} e^b_{~\nu}\,,
\eeqn
holds. The latter is equivalent to imposing the existence of the square-root matrix 
$\sqrt{g^{-1}f}$~\cite{Deffayet:2012zc} which appears in (\ref{effmetr}) as well as in 
the interaction potential of bimetric theory. However, in bimetric theory in 
vierbein formulation with matter coupled to the metric $\tilde{G}_{\mu\nu}$, the condition
(\ref{symcond}) is incompatible with the equations of motion~\cite{Hinterbichler:2015yaa}.
Hence, the two couplings cannot be made equivalent. 
This implies in particular that the vierbein theory
with effective matter coupling does not possess a formulation in terms of metrics.

The two effective matter couplings above have been extensively studied in the literature 
and their phenomenology has already been widely explored in the context of cosmology
(see, e.g., \cite{Enander:2014xga, Comelli:2015pua, Gumrukcuoglu:2015nua} for early works).
The effective theory avoids the ghost at low energies but at high energies it is not consistent 
and requires a ghost-free completion.
Finding such a completion is of particular interest because the effective
metrics have the interesting property that they can couple the massless spin-2 mode 
alone to matter~\cite{Schmidt-May:2014xla}.

The aim of the present work is to construct a symmetric coupling for the two tensor fields $\gmn$ and $\fmn$ 
of bimetric theory to the same matter source, keeping the theory free from the Boulware-Deser ghost even at high energies. 
We obtain this matter coupling by integrating out a non-dymamical field in ghost-free {\bf tri}metric theory.
For low energies our result reduces to the known coupling through the effective metric~(\ref{effmetrvb}). 
At high energies, the coupling in the bimetric setup is highly nontrivial. 
In particular, it does not possess the same form as in GR. Nevertheless, it is always possible to express the 
theory in a simple way (and in terms of a GR coupling) using the trimetric action, 
which essentially provides a formulation in terms of auxiliary fields.

%%%%%%%%%%%%%%%%%%%%%%%%%%%%%%%%%%%%%%%%%%%%%%%%
\section{Ghost-free trimetric theory}
%%%%%%%%%%%%%%%%%%%%%%%%%%%%%%%%%%%%%%%%%%%%%%%%

%%%%%%%%%%%%%%%%%%%%%%%%%%%%%%%%%%%%%%%%%%%%%%%%
\subsection{Trimetric action}
%%%%%%%%%%%%%%%%%%%%%%%%%%%%%%%%%%%%%%%%%%%%%%%%
We will work with the following ghost-free trimetric action for the 
three symmetric tensor fields $\gmn$, $\fmn$ and $\hmn$,
\beqn\label{trimact}
S[g,f,h]=
S_\mathrm{EH}[g]+S_\mathrm{EH}[f]+S_\mathrm{EH}[h]
+S_\mathrm{int}[h,g]
+S_\mathrm{int}[h,f]
+\epsilon S_\mathrm{matter}[h, \phi_i]\,.
\eeqn
It includes the Einstein-Hilbert terms,
\beqn
S_\mathrm{EH}[g]=m_g^2\int\dd^4x~\sqrt{g}~R(g)\,,
\eeqn
with ``Planck mass" $m_g$ and the bimetric interactions,
\beqn\label{intact}
S_\mathrm{int}[h,g]=-2\int\dd^4x~\sqrt{h}~\sum_{n=0}^4\beta^{g}_n\,e_n\big(\sqrt{h^{-1}g}\big)\,,
\eeqn
with parameters $\beta^g_n$ (and $\beta^f_n$ for $S_\mathrm{int}[h,f]$). 
In our parameterization these interaction parameters 
carry mass dimension 4. The scalar functions $e_n$ are the elementary symmetric
polynomials, whose general form will not be relevant in the following.
For later use, we only note that they satisfy,
\beqn\label{deten}
\det (\mathbb{1}+X)=\sum_{n=0}^4e_n(X)\,,
\qquad
e_n(\lambda X)=\lambda^n e_n(X)\,,~~\lambda\in\mathbb{R}\,.
\eeqn
$S_\mathrm{matter}[h, \phi_i]$ is a standard matter coupling (identical to the one in GR) 
for the metric $\hmn$.\footnote{Throughout the whole paper we will use a notation for the matter
action which suggests that the source contains only bosons.
For fermions, it is the vierbein of $\hmn$ that appears in the matter coupling. However,
since we will anyway work in the vierbein formulation later on, this is not a problem and the 
matter coupling to fermions is also covered by our analysis.}
For later convenience, we have included it in the action with a dimensionless parameter $\epsilon$ in front.
As already mentioned in the introduction, consistency does not allow
the other two metrics to couple to the same matter sector.

The structure of the action is dictated by the absence of the 
Boulware-Deser ghost. At this stage, (\ref{trimact}) is the most general trimetric theory 
known to be free from this instability.
In particular, the interactions between the three metrics can only be pairwise through the 
above bimetric potentials and must not form any loops~\cite{Hinterbichler:2012cn, Nomura:2012xr}. 
Moreover, they only contain five free parameters each
and are functions of the square-root matrices $\sqrt{h^{-1}g}$ and $\sqrt{h^{-1}f}$.
The existence of real square-root matrices 
is in general not guaranteed and needs to be imposed on the theory as additional 
constraints for the action to be well-defined. 
At the same time, these constraints ensure a compatible causal structure of the two
metrics under the square root~\cite{Hassan:2017ugh}.

In this paper we will focus on a particular model with 
$\beta^{g}_n=\beta^{f}_n=0$ for $n\geq 2$ in the limit
$m^2_h\rightarrow0$. The choice of interaction parameters significantly 
simplifies the equations and the limit makes the field $\hmn$ non-dynamical.
The potential in this case simply reads,
\beqn\label{intact2}
S_\mathrm{int}[h,g]=-2\int\dd^4x~\sqrt{h}~\Big(\beta^{g}_0+\beta^{g}_1\Tr\sqrt{h^{-1}g}\Big)\,,
\eeqn 
and similar for $S_\mathrm{int}[h,f]$.

%%%%%%%%%%%%%%%%%%%%%%%%%%%%%%%%%%%%%%%%%%%%%%%%
\subsection{Vierbein formulation}\label{sec:vb}
%%%%%%%%%%%%%%%%%%%%%%%%%%%%%%%%%%%%%%%%%%%%%%%%
It will become necessary later on to work in the vierbein formulation
first introduced in~\cite{Hinterbichler:2012cn}.
Therefore we define the vierbeine for the three metrics,
\beqn\label{defvb}
\gmn=e^a_{~\mu}\eta_{ab} e^b_{~\nu}\,,\qquad
\fmn=v^a_{~\mu}\eta_{ab} v^b_{~\nu}\,,\qquad
\hmn=u^a_{~\mu}\eta_{ab} u^b_{~\nu}\,.
\eeqn
Existence of the square-root matrices in the interaction 
potential requires them to satisfy the following symmetry 
constraints~\cite{Hinterbichler:2012cn, Deffayet:2012zc},
\beqn\label{symconstr}
e^\mathrm{T}\eta u=u^\mathrm{T}\eta e\,,\qquad
v^\mathrm{T}\eta u=u^\mathrm{T}\eta v\,,
\eeqn
which we have expressed using matrix notation.
When they are imposed the square-roots can be evaluated to give,
\beqn
\sqrt{h^{-1}g}=u^{-1}e\,,\qquad
\sqrt{h^{-1}f}=u^{-1}v\,.
\eeqn
The interaction potential in $S_\mathrm{int}[h,g]+S_\mathrm{int}[h,f]=-\int\dd^4x\,V$ 
can then be written in the form,
\beqn\label{potvb}
V(e,v,u)=2(\det u)~\Big(\beta^{g}_0+\beta^{f}_0+\beta^{g}_1\Tr[u^{-1}e]+\beta^{f}_1\Tr[u^{-1}v]\Big)\,.
\eeqn
In our particular trimetric model, the constraints (\ref{symconstr}) follow dynamically
from the equations of motion for $e$ and $v$, which was already noticed
in Ref.~\cite{Hinterbichler:2012cn, Deffayet:2012zc}.
We review the underlying argument in a bit more detail because 
it will become relevant for our analysis later.
Namely, the equations for $e$ contain six constraints
arising from local Lorentz symmetry.
In order to make this more precise, we split up the Lagrangian 
$\mathcal{L}=\mathcal{L}_\mathrm{sep}+\mathcal{L}_\mathrm{sim}$ 
into terms $\mathcal{L}_\mathrm{sep}$ that are invariant 
under \textit{separate} Lorentz transformations and $\mathcal{L}_\mathrm{sim}$ that are only invariant under 
\textit{simultaneous} Lorentz transformations of the three vierbeine. 
Their invariance under separate linearized Lorentz transformations of $e$
can be used to show that the terms $\mathcal{L}_\mathrm{sep}$ satisfy the identity,
\beqn\label{idlor}
\frac{\delta \mathcal{L}_\mathrm{sep}}{\delta e}\,\eta^{-1} \,(e^{-1})^\mathrm{T}
-e^{-1}\eta^{-1} \left(\frac{\delta \mathcal{L}_\mathrm{sep}}{\delta e}\right)^\mathrm{T}=0
\,.
\eeqn
The equations of motion 
$\frac{\delta \mathcal{L}_\mathrm{sep}}{\delta e}+\frac{\delta \mathcal{L}_\mathrm{sim}}{\delta e}=0$ 
then imply that the remaining terms $\mathcal{L}_\mathrm{sim}$ in the action will be constrained 
to satisfy~(\ref{idlor}) on-shell,
\beqn\label{constlor}
\frac{\delta \mathcal{L}_\mathrm{sim}}{\delta e}\,\eta^{-1} \,(e^{-1})^\mathrm{T}
-e^{-1}\eta^{-1} \left(\frac{\delta \mathcal{L}_\mathrm{sim}}{\delta e}\right)^\mathrm{T}=0\,.
\eeqn
Using the same arguments, we get a similar constraint for $v$,\footnote{Due to one overall Lorentz
invariance of the action, the constraint obtained from the equations for $u$ will be equivalent to
(\ref{constlor}) and (\ref{constlorv}).}
\beqn\label{constlorv}
\frac{\delta \mathcal{L}_\mathrm{sim}}{\delta v}\,\eta^{-1} \,(v^{-1})^\mathrm{T}
-v^{-1}\eta^{-1} \left(\frac{\delta \mathcal{L}_\mathrm{sim}}{\delta v}\right)^\mathrm{T}=0\,.
\eeqn
Finally, with $\mathcal{L}_\mathrm{sim}=-\int\dd^4 x ~V(e,v,u)$ and (\ref{potvb}), it is straightforward to 
show that (\ref{constlor}) and (\ref{constlorv}) imply the symmetry of $u^{-1}\eta^{-1} (e^{-1})^\mathrm{T}$
and $u^{-1}\eta^{-1} (v^{-1})^\mathrm{T}$, which is equivalent to the constraints (\ref{symconstr}).\footnote{The
last statement follows trivially from $\mathbb{1}=\mathbb{1}^\mathrm{T}=(SS^{-1})^\mathrm{T}
=(S^{-1})^\mathrm{T}S^\mathrm{T}=(S^{-1})^\mathrm{T}S$ for any symmetric matrix $S$.}

%%%%%%%%%%%%%%%%%%%%%%%%%%%%%%%%%%%%%%%%%%%%%%%%
\subsection{Equations of motion}
%%%%%%%%%%%%%%%%%%%%%%%%%%%%%%%%%%%%%%%%%%%%%%%%
From now on we focus on the limit $m^2_h\rightarrow0$ which freezes out the dynamics
of the metric $\hmn$ by removing its kinetic term $S_\mathrm{EH}[h]$ from the action.
In this limit we can solve the equation of motion for $\hmn$ (or its vierbein $u^a_{~\mu}$) algebraically 
and integrate out the nondynamical field. The trimetric action hence assumes the form
of a bimetric theory augmented by an auxiliary field.\footnote{Note that this limit is 
conceptually different from the ones studied in the context of bimetric theory in earlier
works~\cite{Baccetti:2012bk, Hassan:2014vja} since it freezes out the metric that is coupled to the matter sector.} 
Technically, it would be sufficient to assume that $m_h$ 
is negligible compared to all other relevant energy scales in the theory 
(the two other Planck masses, the spin-2 masses and the energies of matter particles).
All our findings can thus also be thought of as being a zeroth-order approximation to trimetric 
theory with very tiny values for $m_h\neq 0$.

For $m^2_h=0$ the equations of motion obtained by varying the action~(\ref{intact2}) 
with respect to the inverse vierbein $u_a^{~\mu}$ are~\cite{Hassan:2011vm},
\beqn\label{withmatter}
\beta_1^g  e^a_{~\mu}+\beta_1^f  v^a_{~\mu}
-\Big(\beta^{g}_0+\beta^{f}_0+\beta^{g}_1\Tr[u^{-1}e]+\beta^{f}_1\Tr[u^{-1}v]\Big)u^a_{~\mu} 
=-\epsilon\, T^a_{~\mu}\,,
\eeqn
where we have introduced the ``vierbein" stress-energy tensor,
\beqn
T^a_{~\mu}\equiv T^a_{~\mu}(u,\phi_i)\equiv-\frac{1}{2\det u}\frac{\delta S_\mathrm{matter}}{\delta u_a^{~\mu}}\,.
\eeqn
It will be easier to work with a form of the equations without the traces appearing.
Tracing equation (\ref{withmatter}) with $u_a^{~\mu}$ gives,
\beqn
\beta^{g}_1\Tr[u^{-1}e]+\beta^{f}_1\Tr[u^{-1}v]
=-\frac{4(\beta^g_0+\beta^f_0)}{3}+\frac{\epsilon}{3} u_a^{~\mu}T^a_{~\mu}\,.
\eeqn
We insert this into (\ref{withmatter}) and obtain,
\beqn\label{withmatter2}
\beta_1^g  e^a_{~\mu}+\beta_1^f  v^a_{~\mu}+\frac{\beta^g_0+\beta^f_0}{3}u^a_{~\mu}
=\epsilon\mathcal{T}^a_{~\mu}\,,
\eeqn
with,
\beqn
\mathcal{T}^a_{~\mu}=\mathcal{T}^a_{~\nu}(u,\phi_i)\equiv 
\frac{1}{3}u^a_{~\mu} u_b^{~\rho}T^b_{~\rho}-T^a_{~\mu}\,.
\eeqn
Our aim in the following is to solve equation (\ref{withmatter2}) for $u^a_{~\mu}$,
plug back the solution into the trimetric action and interpret the result
as an effective bimetric theory with modified matter coupling.

%%%%%%%%%%%%%%%%%%%%%%%%%%%%%%%%%%%%%%%%%%%%%%%%
\section{Vacuum solutions}
%%%%%%%%%%%%%%%%%%%%%%%%%%%%%%%%%%%%%%%%%%%%%%%%

%%%%%%%%%%%%%%%%%%%%%%%%%%%%%%%%%%%%%%%%%%%%%%%%
\subsection{Exact solution for $\hmn$}
%%%%%%%%%%%%%%%%%%%%%%%%%%%%%%%%%%%%%%%%%%%%%%%%

In vacuum with $\epsilon=0$, equation (\ref{withmatter2})  
straightforwardly gives the solution for the vierbein $u$ in terms of $e$ and $v$.
In matrix notation it reads,
\beqn\label{usol}
u=-\frac{3}{\beta^g_0+\beta^f_0}\Big(\beta_1^ge +\beta_1^fv\Big)\,.
\eeqn
The corresponding expression for the metric is,
\beqn\label{hsolsc}
h=u^\mathrm{T}\eta u
=
\Big( ae +bv\Big)^\mathrm{T}\eta\Big(ae +bv\Big)\,,
\eeqn
with constants,
\beqn\label{defab}
a\equiv\frac{3\beta_1^g}{\beta^g_0+\beta^f_0}\,,\qquad
b\equiv\frac{3\beta_1^f}{\beta^g_0+\beta^f_0}\,.
\eeqn
The solution (\ref{hsolsc}) has the same form as the effective metric (\ref{effmetrvb}).

The additional symmetrization constraint $e^\mathrm{T}\eta v=v^\mathrm{T}\eta e$
is equivalent to the existence of the square-root matrix $\sqrt{g^{-1}f}$. 
But, in general, it is not obvious that the existence of this matrix is 
automatically guaranteed by the existence of both $\sqrt{h^{-1}g}$ and 
$\sqrt{h^{-1}f}$. However, in our setup, the symmetrization constraint
is ensured to be satisfied dynamically.
To see this, we simply insert the solution (\ref{usol}) for $u$ into one of the
dynamical trimetric constraints (\ref{symconstr}). This gives,
\beqn\label{symconstr2}
0&=&e^\mathrm{T}\eta u-u^\mathrm{T}\eta e\nn\\
&=&\frac{3}{\beta^g_0+\beta^f_0}\left[\big(\beta_1^ge +\beta_1^fv\big)^\mathrm{T}
\eta e-e^\mathrm{T}\eta \big(\beta_1^ge +\beta_1^fv\big)\right]
=\frac{3\beta_1^f}{\beta^g_0+\beta^f_0}\left[v^\mathrm{T}\eta e-e^\mathrm{T}\eta v\right]\,,
\eeqn
which thus directly implies,
\beqn\label{constraint}
e^\mathrm{T}\eta v-v^\mathrm{T}\eta e=0\,.
\eeqn
The fact that $e^\mathrm{T}\eta v$ is guaranteed to be symmetric dynamically
will become important in the following.

As already stated in the introduction, when (\ref{constraint}) holds,
we can write the right-hand side in terms of metrics,
\beqn\label{solh}
h=ga^2+2ab\, g\big(\sqrt{g^{-1}f}\,\big)+b^2f\,.
\eeqn
The solution for $\hmn$ thus also coincides with the effective metric (\ref{effmetr}).

%%%%%%%%%%%%%%%%%%%%%%%%%%%%%%%%%%%%%%%%%%%%%%%%%%%%%%%%%%%%%%%%%%%%%%%%
\subsection{Effective bimetric potential}\label{sec:effpot}
%%%%%%%%%%%%%%%%%%%%%%%%%%%%%%%%%%%%%%%%%%%%%%%%%%%%%%%%%%%%%%%%%%%%%%%%

We now compute the effective potential for the two dynamical vierbeine.
To this end, we insert the solution (\ref{usol}) for $u$ 
%(for now without imposing $e^\mathrm{T}\eta v=v^\mathrm{T}\eta e$) 
into the trimetric potential (\ref{potvb}).
This gives the effective potential,
\beqn\label{veffvac}
V_\mathrm{eff}(e,v)
=-\frac{54}{\big(\beta^{g}_0+\beta^{f}_0\big)^3}\det \Big(\beta_1^ge +\beta_1^fv\Big)
=\det e\sum_{n=0}^4\beta_n e_n\big(e^{-1}v\big)
\,,
\eeqn
with interaction parameters,
\beqn\label{betan}
\beta_n\equiv B \left(\frac{\beta_1^f}{\beta_1^g}\right)^n\,,
\qquad 
B\equiv-\frac{54(\beta_1^g)^4}{\big(\beta^{g}_0+\beta^{f}_0\big)^3}\,.
\eeqn
In the second equality of (\ref{veffvac}) we have used (\ref{deten}).

The vacuum action for $e$ and $v$ with potential (\ref{veffvac}) is consistent if and only if 
the symmetrization constraint $e^\mathrm{T}\eta v=v^\mathrm{T}\eta e$ holds~\cite{deRham:2015cha}. 
This is a crucial point: An inconsistent theory without this constraint in vacuum
should not arise from our consistent trimetric setup.
The issue gets resolved because the constraint 
is implied by the equations of motion, as we saw in the previous subsection.
Invoking this symmetry constraint we can replace 
$e^{-1}v=\sqrt{g^{-1}f}$ in (\ref{veffvac}) which gives back a ghost-free bimetric theory
with $\beta_n$ parameters given as in~(\ref{betan}).
In conclusion, the effective theory obtained by integrating out $\hmn$ in vacuum 
is identical to a ghost-free bimetric theory. 

Of course, it must also be possible to obtain the constraint (\ref{constraint}) 
in the effective theory, i.e.~without using the equations for $u$ derived in the trimetric setup.
We will verify this in the following by revisiting the arguments given at the end of section~\ref{sec:vb}.
In the present case, the Einstein-Hilbert kinetic terms for $e$ and $v$ 
belong to $\mathcal{L}_\mathrm{sep}$ while $\mathcal{L}_\mathrm{sim}=-V_\mathrm{eff}$.
Thus the constraints arising from the equations of motions after using 
the identity~(\ref{idlor}) set the antisymmetric part of  
$\frac{\delta }{\delta e}V_\mathrm{eff}\eta^{-1} (e^{-1})^\mathrm{T}$ to zero,
\beqn\label{consteq}
\frac{\delta V_\mathrm{eff}}{\delta e}\,\eta^{-1} \,(e^{-1})^\mathrm{T}
-e^{-1}\eta^{-1}\left(\frac{\delta V_\mathrm{eff}}{\delta e}\right)^\mathrm{T}=0\,.
\eeqn
We now solve this constraint explicitly. The variation of (\ref{veffvac}) with respect to $e$ is,
\beqn
\frac{\delta V_\mathrm{eff}}{\delta e}=
B\det \left(e + \frac{\beta_1^f}{\beta_1^g}v\right) \left(e + \frac{\beta_1^f}{\beta_1^g}v\right)^{-1}\,.
\eeqn
We thus have that,
\beqn\label{vareveff}
\frac{\delta V_\mathrm{eff}}{\delta e}\,\eta^{-1} \,(e^{-1})^\mathrm{T}
= B\det \left(e + \frac{\beta_1^f}{\beta_1^g}v\right) 
\left(e + \frac{\beta_1^f}{\beta_1^g} v\right)^{-1}\eta^{-1} \,(e^{-1})^\mathrm{T}\,.
\eeqn
The expression on the right-hand side is a matrix with two upper coordinate indices 
which is constrained to be symmetric by (\ref{consteq}). But this implies that
also its inverse must be symmetric. The inverse of (\ref{vareveff}) is,
\beqn
e^\mathrm{T}\eta\left(\frac{\delta V_\mathrm{eff}}{\delta e}\right)^{-1}
=  B^{-1}\det \left(e + \frac{\beta_1^f}{\beta_1^g}v\right)^{-1} 
e^\mathrm{T}\eta\left(e + \frac{\beta_1^f}{\beta_1^g} v\right)
\eeqn
whose antisymmetric part is precisely 
proportional to $(e^\mathrm{T}\eta v-v^\mathrm{T}\eta e)$. The latter hence vanishes dynamically
and we re-obtain (\ref{constraint}).

The symmetrization constraint remains the same if one couples matter to $e$ or $v$ alone because this 
coupling is invariant under separate Lorentz transformations and therefore does not contribute to
equation (\ref{consteq}). More general matter coupling involving both $e$ and $v$ are only invariant under
simultaneous Lorentz transformations of the vierbeine and thus give rise to extra terms in (\ref{consteq}). 
We will encounter such a situation below.

%%%%%%%%%%%%%%%%%%%%%%%%%%%%%%%%%%%%%%%%%%%%%%%%%%%%%%%%%%%%%%%%%%%%%%%%%%%%%%%%%%%%%%%%%%%%%%
\section{Perturbative solution in the presence of matter}\label{sec:pert}
%%%%%%%%%%%%%%%%%%%%%%%%%%%%%%%%%%%%%%%%%%%%%%%%%%%%%%%%%%%%%%%%%%%%%%%%%%%%%%%%%%%%%%%%%%%%%%
%%%%%%%%%%%%%%%%%%%%%%%%%%%%%%%%%%%%%%%%%%%%%%%%%%%%%%%%%%%%%%%%%
\subsection{Solution for $\hmn$}
%%%%%%%%%%%%%%%%%%%%%%%%%%%%%%%%%%%%%%%%%%%%%%%%%%%%%%%%%%%%%%%%%

In order to derive the solution for the nondynamical field in the presence of a matter source, 
we again work in the vierbein formulation with $e$, $v$ and $u$ defined as in (\ref{defvb})
and with the constraints (\ref{symconstr}) imposed. 

For $\epsilon>0$ we now solve the full equation (\ref{withmatter2}),
\beqn\label{withmatter3}
\beta_1^g  e^a_{~\mu}+\beta_1^f  v^a_{~\mu}+\frac{\beta^g_0+\beta^f_0}{3}u^a_{~\mu}
=\epsilon\mathcal{T}^a_{~\mu}\,,
\eeqn
We can rewrite this in the form (again switching to matrix notation),
\beqn\label{usolmat}
u=\frac{3}{\beta^g_0+\beta^f_0}\Big(\epsilon\mathcal{T}-\beta_1^ge -\beta_1^fv\Big)\,.
\eeqn
Note that, unlike in the vacuum case, this form does now not allow us to express $u$ in terms of
$e$ and $v$ directly, since $u$ still appears on the right-hand side in the stress-energy tensor.
Nevertheless, we can now solve the equations perturbatively.

From now on we shall assume $\epsilon\ll1$, in which case the matter source
can be treated as a small perturbation to the vacuum equations. This allows us to
obtain the solution for $u$ and $h$ as a perturbation series in $\epsilon$. 
To this end, we make the ansatz,
\beqn\label{pertsol}
u=\sum_{n=0}^\infty \epsilon^n u^{(n)}=u^{(0)}+\epsilon u^{(1)}+\mathcal{O}(\epsilon^2)\,,\qquad
h=\sum_{n=0}^\infty \epsilon^n h^{(n)}=h^{(0)}+\epsilon h^{(1)}+\mathcal{O}(\epsilon^2)\,.
\eeqn
The lowest order of the solution is obtained from (\ref{usolmat}) with $\epsilon=0$,
\beqn\label{lou}
u^{(0)}=-\frac{3}{\beta^g_0+\beta^f_0}\Big(\beta_1^ge +\beta_1^fv\Big)\,,
\eeqn
which of course coincides with the solution obtained in vacuum, c.f.~equation (\ref{usol}).
Then the corresponding lowest order in the metric $\hmn$ 
is also the same as in equation (\ref{hsolsc}),
\beqn\label{hsolmat}
h^{(0)}=\big(u^{(0)}\big)^\mathrm{T}\eta u^{(0)}
=\frac{9}{(\beta^g_0+\beta^f_0)^2}
\Big(\beta_1^ge +\beta_1^fv\Big)^\mathrm{T}\eta\Big(\beta_1^ge +\beta_1^fv\Big)\,.
\eeqn
In order to re-arrive at the form (\ref{solh}) in terms of metrics alone we would again have to invoke the 
symmetrization constraint $e^\mathrm{T}\eta v=v^\mathrm{T}\eta e$ which is enforced dynamically 
only at lowest order in perturbation theory. At higher orders, it will be replaced by a new constraint that
needs to be re-computed from the final effective action. Hence, $h^{(0)}$ coincides with the effective metric
(\ref{effmetr}) up to corrections of order $\epsilon$ (which are thus shifted into $h^{(1)}$).

The solutions for the higher orders $u^{(n)}$ in the expansion~(\ref{pertsol}) is given by,
\beqn\label{fullsolu}
u^{(n)}
&=&\frac{3}{\beta^g_0+\beta^f_0}\left.\frac{\delta^n}{\delta\epsilon^n}\Big(\epsilon\mathcal{T}(u, \phi_i)
-\beta_1^ge -\beta_1^fv\Big)\right|_{\epsilon=0}\,,
\eeqn
where in $\mathcal{T}(u,\phi_i)$ one needs to replace $u=\sum_{l=0}^{n-1}\epsilon^l u^{(l)}$, using
the lower-order solutions and further expand in $\epsilon$.
In other words, we can solve for $u^{(n)}$ recursively, using the already 
constructed solutions up to $u^{(n-1)}$.

For instance, the next order $u^{(1)}$ is obtained from (\ref{fullsolu}) 
with $u$ in the stress-energy tensor replaced by $u^{(0)}$,
which gives,
\beqn\label{usolfo}
u^{(1)}=\frac{3}{\beta^g_0+\beta^f_0}\mathcal{T}(u^{(0)},\phi_i)
\,.
\eeqn
The corresponding next order in the metric is therefore,
\beqn\label{hsolfo}
h^{(1)}
&=&\big(u^{(0)}\big)^\mathrm{T}\eta u^{(1)}+\big(u^{(1)}\big)^\mathrm{T}\eta u^{(0)}\nn\\
&=&-\frac{9}{\big(\beta^g_0+\beta^f_0\big)^2}\,
\big(u^{(0)}\big)^\mathrm{T}\eta \mathcal{T}(u^{(0)},\phi_i)+\big(\mathcal{T}(u^{(0)},\phi_i)\big)^\mathrm{T}\eta u^{(0)}\,.
\eeqn
The explicit derivation of the next order requires 
making an assumption for the precise form of the matter source.
Since the solution for $u^{(1)}$ is sufficient to write down the first correction
to the effective action in vacuum, we stop here.

%%%%%%%%%%%%%%%%%%%%%%%%%%%%%%%%%%%%%%%%%%%%%%%%%%%%%%%%%%%%%%%%%
\subsection{Effective action}
%%%%%%%%%%%%%%%%%%%%%%%%%%%%%%%%%%%%%%%%%%%%%%%%%%%%%%%%%%%%%%%%%

Plugging back the solutions for $u$ (or $h$) into the action with potential 
(\ref{potvb}) results in an effective bimetric theory,
perturbatively expanded in $\epsilon$ and written in terms of vierbeine,
\beqn
S_\mathrm{eff}&=& S_\mathrm{EH}[g]+S_\mathrm{EH}[f]\nn\\
&~&-2\int\dd^4x~\Big(\det u\Big)~\Big(\beta^{g}_0+\beta^{f}_0+\beta^{g}_1\Tr[u^{-1}e]+\beta^{f}_1\Tr[u^{-1}v]\Big)
+\epsilon S_\mathrm{matter}[h, \phi_i]\,,
\eeqn
with $u=\sum_{n=0}^\infty\epsilon^nu^{(n)}$ and $u^{(n)}$ given by (\ref{fullsolu}).
Expanding in $\epsilon$, we find that the lowest order terms read,
\beqn
S_\mathrm{eff}
&=& S_\mathrm{EH}[g]+S_\mathrm{EH}[f]\nn\\
&-&2\int\dd^4x~\Big(\det u^{(0)}\Big)\Big(1+\epsilon\,\Tr \Big[(u^{(0)})^{-1}u^{(1)}\Big]\Big)~\Big(\beta^{g}_0+\beta^{f}_0
+\Tr\big[(u^{(0)})^{-1}(\beta^{g}_1e+\beta^{f}_1v)\Big)\nn\\
&+&{2\epsilon}\int\dd^4x\,\Big(\det u^{(0)}\Big)~
\Tr\Big[(u^{(0)})^{-1}u^{(1)}(u^{(0)})^{-1}(\beta^{g}_1e+\beta^{f}_1v)\Big]\nn\\
&+&\epsilon S_\mathrm{matter}[h^{(0)}, \phi_i] 
~+~\mathcal{O}(\epsilon^2)\,.
\eeqn
A short computation shows that, after inserting the expressions 
(\ref{lou}) and (\ref{usolfo}) for $u^{(0)}$ and $u^{(1)}$, this simply becomes,
\beqn\label{effact}
S_\mathrm{eff}
&=&S_\mathrm{EH}[g]+S_\mathrm{EH}[f]+S_\mathrm{int}[e,v]+\epsilon S_\mathrm{matter}[h^{(0)}, \phi_i]
~+~\mathcal{O}(\epsilon^2)\,.
\eeqn
Here, the interaction potential is the one which we already found in section~\ref{sec:effpot},
\beqn\label{effactpot}
S_\mathrm{int}[e,v]\equiv 
-\int\dd^4x~\det e\sum_{n=0}^4\beta_n e_n\big(e^{-1}v\big)\,,
\qquad
\beta_n\equiv -\frac{54(\beta_1^g)^4}{\big(\beta^{g}_0+\beta^{f}_0\big)^3} 
\left(\frac{\beta_1^f}{\beta_1^g}\right)^n\,.
\eeqn
Note that this is not the most general ghost-free bimetric potential since the five $\beta_n$ 
are not independent. They satisfy $\beta_n=\beta_0(\beta_1/\beta_0)^n$ for $n\geq 2$ and hence
the potential in $S_\mathrm{int}[e,v]$ really contains only two free parameters.

Moreover, the effective metric $h^{(0)}$ in the matter coupling is of the form (\ref{effmetrvb})
but the coefficients $a$ and $b$ are not fully independent of the interaction parameters
$\beta_n$ in the potential. More precisely, they satisfy $b/a=\beta_1/\beta_0$.

%%%%%%%%%%%%%%%%%%%%%%%%%%%%%%%%%%%%%%%%%%%%%%%%%%%%%%%%%%%%%%%%%
\subsection{Symmetrization constraints}
%%%%%%%%%%%%%%%%%%%%%%%%%%%%%%%%%%%%%%%%%%%%%%%%%%%%%%%%%%%%%%%%%
The symmetrization constraints~(\ref{symconstr}) in trimetric theory
(which in our model follow from the trimetric equations of motion even in the presence 
of matter) can be treated perturbatively in a straightforward way. 
Using (\ref{pertsol}) we expand them as follows,
\beqn\label{pertcons}
\sum_{n=0}^\infty \epsilon^ne^\mathrm{T}\eta u^{(n)}
=\sum_{n=0}^\infty \epsilon^n(u^{(n)})^\mathrm{T}\eta e\,,\qquad
\sum_{n=0}^\infty \epsilon^nv^\mathrm{T}\eta u^{(n)}
=\sum_{n=0}^\infty \epsilon^n(u^{(n)})^\mathrm{T}\eta v\,.
\eeqn
Comparing orders of the expansion parameter $\epsilon$, we obtain $\forall n$,
\beqn
e^\mathrm{T}\eta u^{(n)}
=(u^{(n)})^\mathrm{T}\eta e\,,\qquad
v^\mathrm{T}\eta u^{(n)}
=(u^{(n)})^\mathrm{T}\eta v\qquad\,.
\eeqn
These constraints on $u^{(n)}$ imply that at each order 
in the perturbation series the square-root matrices exist and we have that,
\beqn
\sqrt{(h^{(n)})^\mathrm{-1}g}=(u^{(n)})^\mathrm{-1}e\,,
\qquad
\sqrt{(h^{(n)})^\mathrm{-1}f}=(u^{(n)})^\mathrm{-1}v\,,
\eeqn
ensuring the perturbative equivalence of the metric and vierbein formulations
in the trimetric theory.

The situation in the effective theory (\ref{effact}) obtained by integration out 
$\hmn$ is more subtle. Namely, the constraint (\ref{constraint}) is obtained dynamically 
only in vacuum. In the presence of matter, it will receive corrections of order
$\epsilon$ and higher. As a consequence, the effective action will in general not 
be expressible in terms of metrics.

The corrections to the vacuum constraint can again be 
straightforwardly obtained by inserting the solution
for the vierbein $u$ into either of the symmetrization constraints in (\ref{pertcons}).
This gives the effective constraint as a perturbation series in $\epsilon$,
\beqn\label{corrsc}
0&=&\tfrac{3}{\beta^g_0+\beta^f_0}\big(e^\mathrm{T}\eta u-u^\mathrm{T}\eta e\big)\nn\\
&=&\beta_1^f\big(v^\mathrm{T}\eta e-e^\mathrm{T}\eta v\big)
+ \epsilon \left[e^\mathrm{T}\eta \mathcal{T}(u^{(0)},\phi_i)
-\left(\mathcal{T}(u^{(0)},\phi_i)\right)^\mathrm{T}\eta e\right]
+\mathcal{O}(\epsilon^2)\,.
\eeqn
In principle, this equation can again be solved recursively and the $\mathcal{O}(\epsilon)$
correction is obtained by using the lowest-order solution 
$v^\mathrm{T}\eta e=e^\mathrm{T}\eta v$ in the terms proportional
to $\epsilon$. It demonstrates that in the effective theory with matter coupling, 
the antisymmetric part of $e^\mathrm{T}\eta v$ is no longer zero but proportional to an 
antisymmetric matrix depending on the matter stress-energy tensor.

%%%%%%%%%%%%%%%%%%%%%%%%%%%%%%%%%%%%%%%%%%%%%%%%
\section{Features of the low-energy theory}
%%%%%%%%%%%%%%%%%%%%%%%%%%%%%%%%%%%%%%%%%%%%%%%%

%%%%%%%%%%%%%%%%%%%%%%%%%%%%%%%%%%%%%%%%%%%%%%%%%%%%%%%
\subsection{Validity of the effective description}
%%%%%%%%%%%%%%%%%%%%%%%%%%%%%%%%%%%%%%%%%%%%%%%%%%%%%%%

Using a specific trimetric setup, we have explicitly constructed a 
ghost-free completion for the effective bimetric action, 
\beqn\label{effact}
S_\mathrm{eff}
&=&S_\mathrm{EH}[g]+S_\mathrm{EH}[f]-\int\dd^4x~\det e\sum_{n=0}^4\beta_n e_n\big(e^{-1}v\big)
+\epsilon S_\mathrm{matter}[\tilde{G}, \phi_i]
%~+~\mathcal{O}(\epsilon^2)
\,,
\eeqn
with matter coupling in terms of the effective metric,
\beqn
\tilde{G}_{\mu\nu}=\big(ae^a_{~\mu} +bv^a_{~\mu}\big)^\mathrm{T}\eta_{ab}\big(ae^b_{~\nu} +bv^b_{~\nu}\big)
\,.
\eeqn
The parameters in (\ref{effact}) obtained in our setup are not all independent but satisfy the relations,
\beqn\label{parconst}
\beta_n=\beta_0(\beta_1/\beta_0)^n\quad \text{for}~n\geq 2\,,
\qquad b/a=\beta_1/\beta_0\,.
\eeqn
The effective description is valid for small energy densities in the matter sector, which we 
have parameterized via $\epsilon\ll 1$. It corresponds precisely to the action 
proposed in Ref.~\cite{Noller:2014sta, Hinterbichler:2015yaa}. For higher energies (where the action 
(\ref{effact}) is known to propagate the Boulware-Deser ghost~\cite{deRham:2015cha}) 
the corrections become important. For these energy regimes, parameterized via
$\epsilon\gtrsim 1$, it is simplest to work in the manifestly ghost-free trimetric 
formulation (\ref{trimact}) with $m_h=0$ 
(for instance, if one wants to derive solutions to the equations in the full theory).

Even though the decoupling limits of bimetric theory with matter coupling to $G_{\mu\nu}$
and the vierbein theory with matter coupling to $\tilde{G}_{\mu\nu}$ are identical~\cite{deRham:2015cha},
the two couplings are not equivalent to first order in $\epsilon$.
Namely, the corrections to the symmetrization constraint $e^\mathrm{T}\eta v=v^\mathrm{T}\eta e$ in (\ref{corrsc}), 
are of $\mathcal{O}(\epsilon)$ and thus equally important as the matter coupling itself.
As a consequence, the effective metric $G_{\mu\nu}$ defined in (\ref{effmetr}) differs from $\tilde{G}_{\mu\nu}$
in (\ref{effact}) at $\mathcal{O}(\epsilon)$. Replacing $\tilde{G}_{\mu\nu}$ by $G_{\mu\nu}$ 
in the matter coupling introduces correction terms of $\mathcal{O}(\epsilon^2)$
which we have anyway suppressed in (\ref{effact}). However, the additional terms coming from (\ref{corrsc})
will show up in the interaction potential, which contains the antisymmetric components 
$(e^\mathrm{T}\eta v-v^\mathrm{T}\eta e)$, and contribute at $\mathcal{O}(\epsilon)$. 
Therefore, even when $\mathcal{O}(\epsilon^2)$ terms are neglected, 
the theory with action (\ref{effact}) is not equivalent to bimetric theory 
with matter coupling via the effective metric~$G_{\mu\nu}$.
 
 This picture is consistent with the results in Ref.~\cite{Hinterbichler:2015yaa}, which 
essentially already discussed the $\mathcal{O}(\epsilon)$ correction
in (\ref{corrsc}) and stated that the vacuum constraint $e^\mathrm{T}\eta v=v^\mathrm{T}\eta e$
cannot be imposed when matter is included via the effective vierbein coupling.

%%%%%%%%%%%%%%%%%%%%%%%%%%%%%%%%%%%%%%%%%%%%%%%%
\subsection{The massless spin-2 mode}
%%%%%%%%%%%%%%%%%%%%%%%%%%%%%%%%%%%%%%%%%%%%%%%%

Interestingly, our interactions parameters in (\ref{effact}) 
subject to the constraints (\ref{parconst}) satisfy,
\beqn\label{condpbg}
\frac{cm_f^2}{m_g^2}(\beta_0+3c\beta_1+3c^2\beta_2+c^3\beta_3)&=&\beta_1+3c\beta_2+3c^2\beta_3+c^3\beta_4\,,
\qquad c\equiv\frac{m_g^2}{m_f^2}\frac{b}{a}\,.
\eeqn
This condition was derived in Ref.~\cite{Schmidt-May:2014xla} to ensure that
proportional background solutions of the form $\bar{f}_{\mu\nu}=c^2\bar{g}_{\mu\nu}$ 
exist in bimetric theory with effective matter coupling through the metric $G_{\mu\nu}$ in (\ref{effmetr}). 
In our case with metric $\tilde{G}_{\mu\nu}$, the proportional backgrounds are only solutions in vacuum since 
the corrections to the symmetrization constraint in (\ref{constraint}) 
are in general not compatible with $\bar{v}^a_{~\mu}=c\bar{e}^a_{~\mu}$. 
This situation is comparable to ordinary bimetric theory with matter coupling via
$\gmn$ or $\fmn$. 

Around the proportional vacuum solutions, 
the massless spin-2 fluctuation is~\cite{Hassan:2012wr},
\beqn\label{masslessfluc}
\delta g+\frac{m_f^2}{m_g^2}\delta f
=\delta e^\mathrm{T}\eta \bar{e} + \bar{e}\eta  \delta e^\mathrm{T}
+ \frac{cm_f^2}{m_g^2}\left(\delta v^\mathrm{T}\eta \bar{e} + \bar{e}\eta  \delta v^\mathrm{T}\right)\,.
\eeqn
Around the same background, our effective metric $\tilde{G}_{\mu\nu}$ 
which couples to matter in the effective action~(\ref{effact}) has fluctuations,
\beqn
\delta\tilde{G}_{\mu\nu}&=&
(a+bc)\Big(\bar{e}^\mathrm{T}\eta(a\delta e+b\delta v)
+(a\delta e+b\delta v)\eta\bar{e}^\mathrm{T}\Big)\nn\\
&=&\left(1+\frac{c^2m_f^2}{m_g^2}\right)\left(\delta e^\mathrm{T}\eta \bar{e} + \bar{e}\eta  \delta e^\mathrm{T}
+ \frac{cm_f^2}{m_g^2}\left(\delta v^\mathrm{T}\eta \bar{e} + \bar{e}\eta  \delta v^\mathrm{T}\right)\right)\,,
\eeqn
where we have used (\ref{condpbg}) in the second equality.
The fluctuations of $\tilde{G}_{\mu\nu}$ are proportional to~(\ref{masslessfluc}) and
thus they are purely massless, without containing contributions from the massive spin-2 
mode.\footnote{The fluctuations of $G_{\mu\nu}$ in (\ref{effmetr}) are also proportional 
to~(\ref{masslessfluc}) when the parameters satisfy (\ref{condpbg})~\cite{Schmidt-May:2014xla}.}
We conclude that in the effective theory with action (\ref{effact}), matter interacts only with
the massless spin-2 mode.

%%%%%%%%%%%%%%%%%%%%%%%%%%%%%%%%%%%%%%%%%%%%%%%%
\section{Summary \& discussion}
%%%%%%%%%%%%%%%%%%%%%%%%%%%%%%%%%%%%%%%%%%%%%%%%

We have presented a trimetric setup which at high energies delivers a ghost-free completion
for a well-studied effective matter coupling in bimetric theory.
Our results suggest that even though both effective metrics ${G}_{\mu\nu}$
and $\tilde{G}_{\mu\nu}$ can be coupled to matter without re-introducing the Boulware-Deser
ghost in the decoupling limit, the vierbein coupling via the latter is probably the preferred choice since 
it can be rendered ghost-free by adding additional terms to the action.
Properties of the theory at high energies (description of the Early Universe, black holes, etc.) 
are easy to study in our trimetric formulation and it would be interesting
to revisit phenomenological investigations that have been carried out in the effective theory.

Our results further demonstrate that the metric $\tilde{G}_{\mu\nu}$ in the matter coupling possesses
massless fluctuations around the maximally symmetric vacuum solutions.  
This is of phenomenological relevance because we expect it to avoid constraints arising from the so-called 
vDVZ discontinuity \cite{vanDam:1970vg, Zakharov:1970cc}, which forces the ratio of Planck 
masses $m_f/m_g$ to be small in bimetric theory with ordinary matter coupling~\cite{Enander:2015kda, Babichev:2016bxi}.
These constraint usually arise at distance scales larger than the Vainshtein radius~\cite{Vainshtein:1972sx},
but in case of matter interacting only with the massless spin-2 mode there is no need to invoke the 
Vainshtein mechanism in order to cure the discontinuity.
By the same argument, linear cosmological perturbations are expected to behave similarly to GR.
Subtleties could arise due to the highly nontrivial symmetrization constraint~(\ref{corrsc}) 
and the phenomenology needs to be worked out in detail to explicitly confirm these expectations.

A generalization of our construction to more than two fields in vacuum is studied in~\cite{Hassan:2018mcw} and 
leads to new consistent multi-vierbein interactions.
It would also be interesting to generalize our setup to other values of interactions parameters
in the trimetric action and include more terms in~(\ref{intact2}). 
For the most general set of parameters, it seems difficult to integrate out the vierbein $u$ for the
non-dynamical metric. There may however be simplifying choices different from~(\ref{intact2}) which
allow us to obtain an effective theory with parameters different from (\ref{parconst}).
It would also be interesting to see whether one can find more general forms for effective metrics in this way. 
Possibly, these could be the metrics identified in~\cite{Heisenberg:2014rka} and we leave these interesting
investigations for future work.

\acknowledgments
We thank Mikica Kocic for useful comments on the draft and are particularly grateful to 
Fawad Hassan for making very valuable suggestions to improve the presentation of our results.
This work is supported by a grant from the Max Planck Society.

%%%%%%%%%%%%%%%%%%%%%%%%%%%%%%%%%%%%%%%%%%%%%%%%%%%%%%%%%%%%%%%%%%%%%%%%%%%%%%%%%%%%%%%%%%%%%%%%%%
%%%%%%%%%%%%%%%%%%%%%%%%%%%%%%%%%%%%%%%%%%%%%%%%%%%%%%%%%%%%%%%%%%%%%%%%%%%%%%%%%%%%%%%%%%%%%%%%%%

%%%%%%%%%%%%%%%%%%%%%%%%%%%%%%%%%%%%%%%%%%%%%%%%%%%%%%%%%%%%%%%%%%%%%%%%%%%%%%%%%%%%%%%%%%%
%%%%%%%%%%%%%%%%%%%%%%%%%%%%%%%%%%%%%%%%%%%%%%%%%%%%%%%%%%%%%%%%%%%%%%%%%%%%%%%%%%%%%%%%%%%

%%%%%%%%%%%%%%%%%%%%%%%%%%%%%%%%%%%%%%%%%%%%%%%%%%%%%%%%%%%%%%%%%%%%%%%%%%%%%%%%%%%%%%%%%%%%%%%%%%%%%%%%%%%%%%%%%%%%%%%%%%%%%%%%%%%%%%%%%%%%%%
%%%%%%%%%%%%%%%%%%%%%%%%%%%%%%%%%%%%%%%%%%%%%%%%%%%%%%%%%%%%%%%%%%%%%%%%%%%%%%%%%%%%%%%%%%%%%%%%%%%%%%%%%%%%%%%%%%%%%%%%%%%%%%%%%%%%%%%%%%%%%%
\end{document}